# Towards low-temperature processing of efficient γ-CsPbI$_3$ perovskite solar cells

Zongbao Zhang,[a,b] Ran Ji,[a,b] Yvonne J. Hofstetter,[a,b] Marielle Deconinck,[a,b] Julius Brunner,[a,b] Yanxiu Li,[a,b] Qingzhi An,[a,b] Yana Vaynzof*[a,b]

a. Center For Advancing Electronics Dresden (CFAED), Technical University of Dresden, Helmholtzstraße 18, 01069, Germany,
   E-mail: yana.vaynzof@tu-dresden.de
b. Leibniz-Institute for Solid State and Materials Research Dresden, Helmholtzstraße 20, 01069 Dresden, Germany

Inorganic cesium lead iodide (CsPbI$_3$) perovskite solar cells (PSCs) have attracted enormous attention due to their excellent thermal stability and optical bandgap (~1.73 eV), well-suited for tandem device applications. However, achieving high-performing photovoltaic devices processed at low temperatures is still challenging. Here we reported a new method to fabricate high-efficiency and stable γ-CsPbI$_3$ PSCs at lower temperatures than was previously possible by introducing the long-chain organic cation salt ethane-1,2-diammonium iodide (EDAI$_2$) and regulating the content of lead acetate (Pb(OAc)$_2$) in the perovskite precursor solution. We find that EDAI$_2$ acts as an intermediate that can promote the formation of γ-CsPbI$_3$, while excess Pb(OAc)$_2$ can further stabilize the γ-phase of CsPbI$_3$ perovskite. Consequently, improved crystallinity and morphology and reduced carrier recombination are observed in the CsPbI$_3$ films fabricated by the new method. By optimizing the hole transport layer of CsPbI$_3$ inverted architecture solar cells, we demonstrate up to 16.6% efficiencies, surpassing previous reports examining γ-CsPbI$_3$ in inverted PSCs. Notably, the encapsulated solar cells maintain 97% of their initial efficiency at room temperature and dim light for 25 days, demonstrating the synergistic effect of EDAI$_2$ and Pb(OAc)$_2$ on stabilizing γ-CsPbI$_3$ PSCs.

**Introduction**

In the past decades, organic-inorganic hybrid perovskite solar cells (PSCs) have made great progress, as illustrated by the fact that their power conversion efficiencies (PCEs) have risen from the initial 3.5% to 25.7%.[1–5] However, state-of-the-art perovskite devices often employ organic cations such as formamidinium (FA) and methylammonium (MA), which exhibit poor chemical and thermal stability, hindering their commercial application. Replacing these volatile organic cations with inorganic ions such as Cs and Ru has the potential to improve the intrinsic stability of perovskite devices.[6,7]

For inorganic cesium lead halide perovskites, cesium lead iodide ($CsPbI_3$) PSCs have attracted particularly widespread attention due to their tolerance to high temperatures and an ideal bandgap of ~1.73 eV, which is very suitable for tandem photovoltaic applications.[8–12] Although the performance of $CsPbI_3$ PSCs has surged to over 20%,[13,14] most high-performance devices are fabricated via the dimethylammonium iodide (DMAI)-assisted method, which limits the options for mass preparation.[8,15–18] Furthermore, the potential presence of organic component DMA in the final $CsPbI_3$ films remains under debate.[19–23] Therefore, it is imperative to develop a plethora of preparation processes to adapt to complex industrial manufacturing and obtain highly phase-pure $CsPbI_3$ thin films. Consequently, You and colleagues reported that high efficiency (PCE=15.7%) α-$CsPbI_3$ PSCs can be fabricated by employing a solvent-controlled growth of perovskite precursors under a dry environment. Moreover, the devices did not show an efficiency drop under continuous light soaking for more than 500 hours.[24] Liu *et al.* reported that careful control over the elemental composition and annealing environment enabled the fabrication of γ-$CsPbI_3$ solar cells with a PCE of 16.3%.[25] Alternatively, additives have also proven to be very effective in modulating the perovskite crystallization dynamics, resulting in the formation of high-quality perovskite films. Ho-Baillie and co-workers adopted a cation exchange growth method for fabricating high-quality γ-$CsPbI_3$ by introducing methylammonium iodide (MAI) additive into the perovskite precursor solution, achieving 14.1% device efficiency.[7] Alternatively, Li. *et al.* developed a mediator-antisolvent strategy in which MAI and phenyl-$C_{61}$-butyric acid methyl ester ($PC_{61}BM$) were used as additives introduced into the chlorobenzene antisolvent used for device fabrication. The method allowed for obtaining high-quality and stable black-phase $CsPbI_3$ perovskite films with solar cell efficiencies reaching up to 16.04%.[26] Very recently, Luo's group demonstrated that adding formamidine acetate (FAAc) into the precursor solution improves the phase purity and electronic quality of γ-$CsPbI_3$ films. As a result, the authors achieved a very high efficiency of over 18%.[6] However, despite the impressive achievements outlined above, a key limitation for their future application lies in the fact that all these methods rely on annealing at high temperatures (~340 °C) in order to ensure rapid volatilization of the organic components, which may inevitably trigger the formation of voids and pinholes in the perovskite film. Importantly, the high-temperature processing required to form $CsPbI_3$ by these methods made it necessary to employ the n-i-p architecture of photovoltaic devices. This architecture, however, exhibits several disadvantages, such as a poor thermal stability of the hole transport layers (e.g., Spiro-MeOTAD) and incompatibility with applications in tandem devices, which predominantly rely on a p-i-n configuration.[17,18] Moreover, as compared to standard architecture n-i-p devices, the p-i-n structured solar cells do not rely on the use of undesired dopants in the carrier transport layer, which has been shown to negatively impact the device's stability and hysteresis. [27,28] Finally, inverted p-i-n devices can be realized entirely via low-temperature processing, which is a key requirement for their commercialization on a large-scale as well as their application on flexible substrates. [4,29–31] Therefore, it is imperative to develop low-temperature (<200 °C) processes to prepare $CsPbI_3$ PSCs based on a p-i-n architecture.

Huang's group reported a facile method to stabilize α-$CsPbI_3$ films at low temperatures (~100 °C) with p-i-n structure via a small amount of sulfobetaine zwitterion addition in perovskite solution. This method endowed the solar cells with a PCE of 11.4%.[32] Alternatively, the evaporation process has also been proven to be a suitable method for fabricating $CsPbI_3$ film at low temperatures. Unold. *et al.* prepared a stable γ-$CsPbI_3$ film at 50 °C by elaborately controlling the composition ratios of CsI and $PbI_2$ during the evaporation process. As a result, the authors obtained an efficiency exceeding 12%.[33] In recent years, , we also reported on the vapor deposition of $CsPbI_3$ via a low-temperature process (~100 °C) and proposed a ternary source (CsI, $PbI_2$, and phenylethylammonium iodide) co-evaporation method to prepare the stable γ-$CsPbI_3$ film, achieving a 15% device performance.[34] Nevertheless, the efficiency of devices fabricated at low temperatures still lags far behind those made at high temperatures.

Therefore, fabricating high-efficiency pure $CsPbI_3$ PSCs at a low temperature (<200 °C) remains challenging. Notably, utilizing alternative, non-halide lead sources opens an opportunity to address this issue. For example, lead acetate ($Pb(OAc)_2$) is intensively used to fabricate high-quality $MAPbI_3$ films via a rapid process that relies on a low annealing temperature – a consequence of an accelerated crystallization process facilitated by the removal of the volatile methylammonium acetate (MAOAc).[35–40] Similarly, $Pb(OAc)_2$ has also been utilized as a cost-effective lead source to replace the traditionally used $PbI_2$ for the fabrication of $FA_xMA_{1-x}PbI_3$, and $FA_{1-x}Cs_xPbI_3$ devices, resulting in high-quality perovskite films and good photovoltaic performance.[38,39,41] Other works explored the use of $Pb(OAc)_2$ as an additive in perovskite precursor solutions in order to improve the morphology and stability of the films.[42–44] Despite these promising results, the use of $Pb(OAc)_2$ in fabricating inorganic perovskites remains rare, and the related film formation mechanisms are poorly understood.

In this work, we propose a low-temperature preparation method for high-quality and stable γ-$CsPbI_3$ films, using the long-chain organic cation ethane-1,2-diammonium iodide ($EDAI_2$) as an additive and replacing $PbI_2$ with a more cost-effective lead source-$Pb(OAc)_2$ in the perovskite precursor solution. Our results indicate that the use of $EDAI_2$ and $Pb(OAc)_2$ results in the formation of $EDA(OAc)_2$, which escapes during the film formation process and promotes the formation of γ-$CsPbI_3$. At the same time, excess $Pb(OAc)_2$ additive can further stabilize the γ-$CsPbI_3$ perovskite. Improved crystallinity, morphology and reduced carrier recombination are demonstrated for films with an optimal $Pb(OAc)_2$ excess. Importantly, by replacing $PbI_2$ with $Pb(OAc)_2$ as the Pb source, we eliminate the need to employ dimethylformamide (DMF) – a highly toxic solvent – and rely only on the significantly less toxic dimethyl sulfoxide (DMSO) for the processing of the $CsPbI_3$ films. By further optimizing the type of hole transport layer (HTL) used for inverted architecture device fabrication, we obtained an efficiency of 16.57%, which surpasses those previously reported for inverted γ-$CsPbI_3$ PSCs.[32–34,45–47] The optimized $CsPbI_3$ devices also showed improved stability, with encapsulated solar cells maintaining 97% of their initial efficiency at room temperature and under dim light illumination for 25 days of aging. These findings provide a new method to fabricate high-efficiency and stable inorganic perovskite solar cells using Pb(OAc)$_2$ and with low-temperature processes.

## Results and Discussion

**Impact of Pb(OAc)$_2$ excess on film formation of CsPbI$_3$**

To investigate the effect of Pb(OAc)$_2$ excess, we fabricated perovskite films using EDAI$_2$, CsI, and Pb(OAc)$_2$ with a stoichiometry ratio of 1:1:X in a DMSO solution (Fig. 1a), in which X denotes the relative proportion of Pb(OAc)$_2$ to the other precursors that were introduced in equant amounts. Test experiments revealed that using a perfectly stoichiometric ratio of 1:1:1 did not lead to the formation of a perovskite, so a series of different X values with varying Pb(OA$_C$)$_2$ excess (e.g., 1.1, 1.15, 1.2, 1.25, and 1.3) were studied. The perovskite films were fabricated in a one-step process and subsequently annealed at 60 °C for 5 min, then 180 °C for another 3 min in dry air, as shown in Fig. 1b. Detailed processing procedures are provided in the Experimental Section.

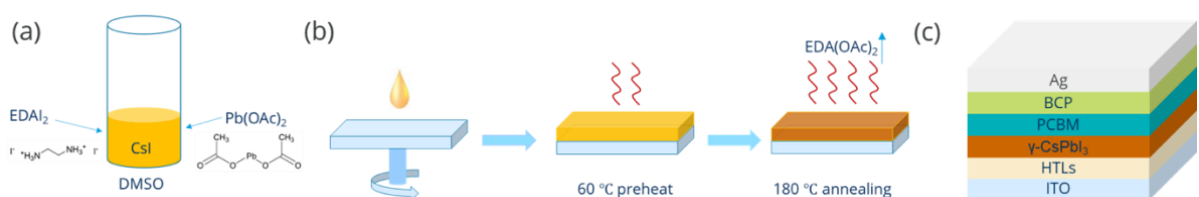

**Fig. 1** Schematic illustration of the precursor preparation (a) and film fabrication procedure (b). (c) Structure of the device in this work.

To examine the evolution of the film formation before and during the annealing process, X-ray diffraction (XRD) measurements for perovskite films undergoing different thermal annealing temperatures and times were carried out. As is shown in Fig. 2a, the as-deposited wet film (prior to annealing) contains DMSO-related intermediate products.[39,48–50] When preheating the film at 60 °C, the intermediate product vanishes, while crystalline PbI$_2$ is formed. As the annealing temperature is increased to 180 °C, the PbI$_2$ is rapidly decreased and already after 30 s can no longer be observed. At the same time, the CsPbI$_3$ and Pb(OAc)$_2$ signals appear. The accompanying images and top-view SEM micrographs of the films at the same annealing conditions are shown in Fig. S1. The images illustrate the evolution of the layer's morphology both in terms of the increase in grain size and the elimination of pinholes from the film's microstructure once the 3 min annealing at 180 °C is completed.

To further understand the process taking place during the preheating, we characterized the preheated film (60 °C, 5 min) using Fourier-transform infrared spectroscopy (FTIR). Fig. 2b shows several pronounced characteristic peaks associated with COO$^-$ and N-H vibrations, which suggests that an EDA(OAc)$_2$ product may form in the preheated film. To further verify whether the EDA(OAc)$_2$ can be removed from the film during the subsequent annealing process (i.e., at 180 °C), we performed thermogravimetric analysis (TGA) for the EDA(OAc)$_2$ intermediate product. As shown in Fig. S2, the initial decomposition temperature (defined by T at 95% weight) for EDA(OAc)$_2$ is 137.7 °C, which confirms that EDA(OAc)$_2$ is thermally unstable and can be easily removed from the perovskite film when the annealing temperature is higher (180 °C). These measurements suggest that the role of EDA(OAc)$_2$ in the formation of CsPbI$_3$ is similar to that of MAAc in the fabrication of MAPbI$_3$ perovskite based on the Pb(OAc)$_2$ recipe[35,36] – in both cases, these species can be removed from the perovskite film during the annealing process. Based on these results, we propose that the crystallization process occurs via the following route:

X·Pb(OAc)$_2$+EDAI$_2$+CsI→CsPbI$_3$+EDA(OAc)$_2$↑+(X-1)·Pb(OAc)$_2$    (X>1)

To characterize the crystalline structure of the CsPbI$_3$ films formed with different amounts of Pb(OAc)$_2$ excess, XRD measurements were carried out, with the results shown in Fig. 2c. It shows that for X>1.1 the diffractograms exhibit the characteristic perovskite peaks at ~14.3° and ~29.0°, corresponding to the (110) and (220) planes of the orthorhombic γ-CsPbI$_3$, respectively. It can be seen that the intensity of the (110) and (220) peaks increases with a rise in the Pb(OAc)$_2$ excess, which suggests that the excessive Pb(OAc)$_2$ facilitates the preferred direction of (110) and (220) in the final perovskite film. For lower Pb(OAc)$_2$ composition (e.g., X=1.1), no (110) and (220) diffraction peaks are observed, which implies that a significant excess of Pb(OAc)$_2$ is a prerequisite for the formation of γ-CsPbI$_3$ perovskite. Examining the diffractograms at low angles (below 12°, Fig. S3a) reveals that a small fraction of the films is δ-phase CsPbI$_3$, with higher X ratios resulting in a small fraction of this unfavorable phase. Notably, the patterns show no characteristic 2D peaks in the final CsPbI$_3$ films, suggesting that EDAI$_2$ is not incorporated into the perovskite lattice. Moreover, for films with a large excess of Pb(OAc)$_2$ (X≥1.25), a diffraction peak at 7.5° can be observed, which corresponds to Pb(OAc)$_2$. To further confirm that no 2D phases are present in the final CsPbI$_3$ film, grazing incidence x-ray diffraction (GIXRD) measurements were carried out on the X=1.2 sample by varying the incidence angle Ω from 0.2 to 2.0 (Fig. S3b). The low-incidence angles are more surface sensitive, while the high angles probe more of the bulk. It can be seen that no characteristic 2D signals are detected from surface to bulk, which means there are no 2D perovskite phases formed in the final film. We note that we observe a very small amount of δ-phase CsPbI$_3$ at the bulk of the film, yet, as will be shown later, this does not seem to impact either the optoelectronic quality or the stability of the perovskite.

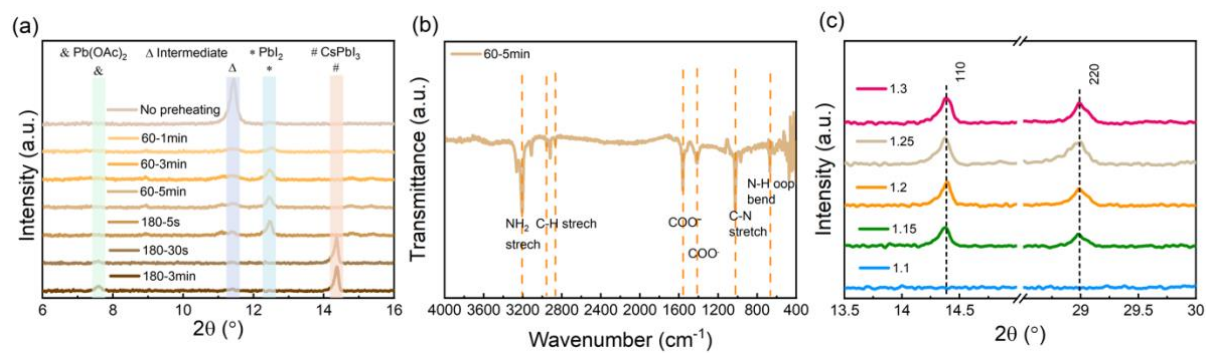

**Fig. 2** (a) X-ray diffraction (XRD) patterns of perovskite film on ITO substrates with different heating temperature and time, (b) Fourier-transform infrared spectroscopy of intermediate perovskite film (after perovskite precursors deposited, the film was preheated at 60 °C for 5 min), (c) XRD patterns of final CsPbI$_3$ on glass substrates with different Pb(OAc)$_2$ excess.

To confirm the absence of EDAI$_2$ in the perovskite layers, we performed X-ray photoemission spectroscopy (XPS) measurements for the preheated and annealing films with different Pb(OAc)$_2$ excess amounts (Fig. S4, Supporting Information). Regardless of the exact amount of Pb(OAc)$_2$ excess, the preheated films show obvious nitrogen signals (corresponding to atomic percentages ranging from 10 to 15%) before the annealing at 180 °C, indicating that the EDA cation is present in the films (Fig. S4). After annealing at 180 °C for 3 minutes, no nitrogen could be detected in the films, confirming that EDA is no longer present. Considering that XPS is a surface-sensitive technique, we performed XPS depth profiling on the 1.2 sample in order to examine whether EDA might be present in the bulk of the film. We observe no nitrogen signals, neither at the surface (Fig. S5) of the layer nor in the bulk (Table S1), which confirms that no EDA cations remain in the final perovskite film. O1s spectrum collected at the surface of the film exhibits two peaks. While the high binding energy peak is likely associated with surface contamination, the low binding energy peak at ~530 eV may arise from either the Pb(OAc)$_2$ precursor or DMSO solvent. Considering that no sulfur signal is observed in the XPS spectra, the latter can be ruled out, suggesting that these oxygen species are associated with the Pb(OAc)$_2$ precursor. We note that the low binding energy peak is unlikely to be related to the formation of lead oxide since it requires much higher temperatures (~350 °C) than those used by us during layer fabrication.[44] The gradual increase in the oxygen and lead atomic concentrations with increasing Pb(OAc)$_2$ ratio confirms that excess Pb(OAc)$_2$ in the precursor solution also results in an excess in the final film, in agreement with the XRD results (Fig. S4).

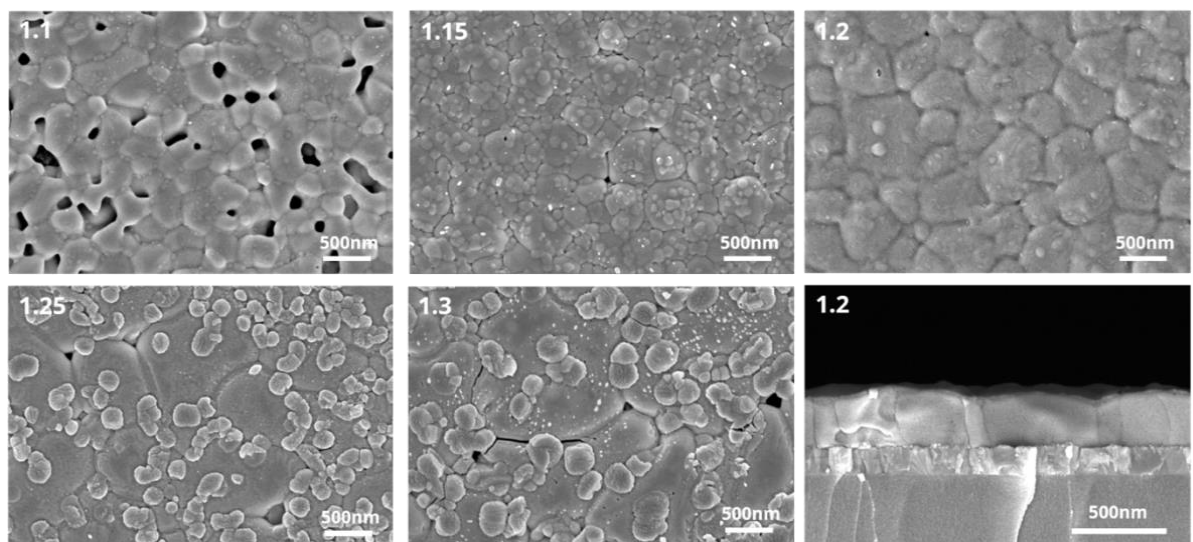

**Fig. 3** Scanning electron microscopy images of different Pb(OAc)$_2$ ratio (X=1.1, 1.15, 1.2, 1.25, 1.3). Cross-sectional images of 1.2 samples on glass/ITO.

To further confirm that no EDA cation was left in the final perovskite film, proton nuclear magnetic resonance (H-NMR) measurements of both the CsPbI$_3$ film and the intermediate product EDA(OAc)$_2$ were carried out, with the results shown in Fig. S6. The EDA(OAc)$_2$ product displays a very strong NMR peak associated with –CH$_2$- ($\delta$=2.7 ppm) and –NH$_3^+$- ($\delta$=5.4 ppm) originating from the EDA$^{2+}$ and a CH$_3$COO$^-$ ($\delta$=1.80 ppm) peak, which is consistent with previous reports in the literature.[51] The H-NMR signals of –CH$_2$- and –NH$_3^+$- disappear entirely in the final CsPbI$_3$ perovskite film, which confirms that no EDA cation remains in the final perovskite layer. We note that there are two additional peaks in the final CsPbI$_3$ film. The peak at ~3.3 ppm corresponds to an impurity such as water in the DMSO-D6 solvent. The weak but non-negligible peak at 1.74 ppm corresponds to CH$_3$COO$^-$. This signal arises from the excess of Pb(OAc)$_2$ present in the film, which agrees with both the XRD and

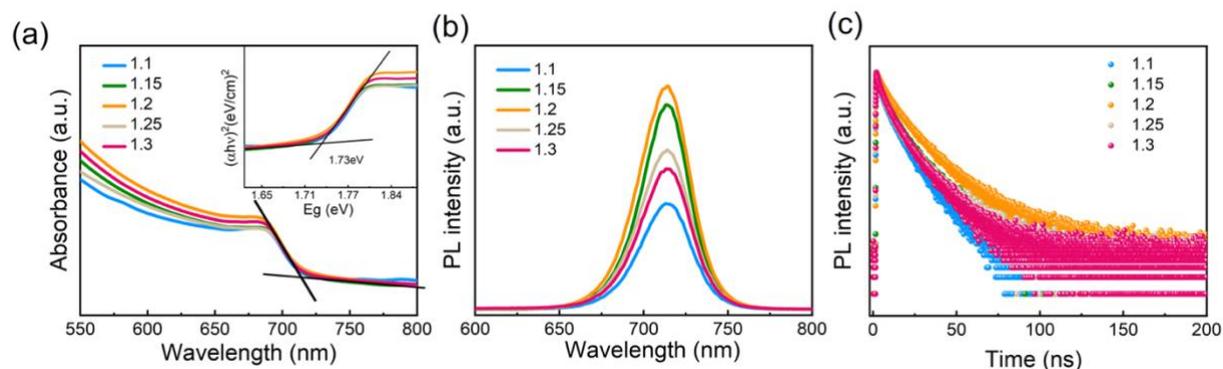

XPS measurements.

**Fig. 4** (a) UV-Vis absorbance spectra (inset: tauc plot for CsPbI$_3$ film with different Pb(OAc)$_2$ excess) and (b) Steady state and (c) time-resolved photoluminescence emission (PL) spectra of different Pb(OAc)$_2$ ratio (1.1, 1.15, 1.2, 1.25, 1.3) samples.

Concerning the crystallinity changes observed for samples with different amounts of Pb(OAc)$_2$ excess, the microstructure of the layers will be expected to play a role as well. To characterize it, we performed scanning electron microscopy (SEM) measurements, and the representative images are shown in Fig. 3. It can be seen that the perovskite films show many obvious pinholes with smaller amounts of Pb(OAc)$_2$ excess. These pinholes gradually disappear as the content of Pb(OAc)$_2$ increases. For X=1.2, no pinholes can be observed and the film consists of relatively large and uniform crystal domains. A further rise in the Pb(OAc)$_2$ excess has a negative impact on the microstructure. We observe the appearance of small additional domains at the surface, which appear different in contrast to the perovskite domains. To examine the composition of these small domains on the surface, energy-dispersive X-ray spectroscopy (EDX) spectra were measured at different sample locations. As is shown in Fig. S7, the small domains exhibit an additional oxygen-related signal which is not present in the large domains associated with the perovskite. These measurements suggest that Pb(OAc)$_2$ remains on the perovskite crystal surface, consistent with the XRD and H-NMR results. To examine the microstructure in the direction of charge transport, cross-sectional SEM was also measured on the X=1.2 sample. Samples with this composition show very compact crystal packing, without voids at the interface to the substrate or in bulk, with many domains protruding throughout the entire layer thickness. This suggests that the X=1.2 stoichiometry enables the formation of high-quality perovskite films with relatively few boundaries in the direction of charge transport.

To investigate the optical properties of the perovskite films with different amounts of Pb(OAc)$_2$ excess, absorption, steady-state and time-resolved photoluminescence (TRPL) measurements were performed. As is shown in Fig. 4a, no significant changes in the UV-vis absorption spectra could be observed. The absorption edges are located at approximately 715 nm, corresponding to a 1.73 eV bandgap, consistent with other reports of the optical properties of γ-CsPbI$_3$ in literature.[10,52,53] PL spectra show no shift in the peak position for the samples with different Pb(OAc)$_2$ compositions, which agrees with the absence of bandgap variation in the absorption spectra. The PL intensity gradually increases with X rising with the maximum achieved at X=1,2, suggesting that the presence of a certain fraction of excess Pb(OAc)$_2$ can passivate the defects in the perovskite bulk or at the domain boundaries, thus reducing the degree of non-radiative recombination.[44,54] However, with further increase in the Pb(OAc)$_2$ excess, the PL intensity is reduced, which we associate with the poorer film microstructure and the presence of Pb(OAc)$_2$ domains.[44,55] Fig 4c displays TRPL decay curves for different amounts of Pb(OAc)$_2$ excess. Considering that the curves show a non-monoexponential decay, the lifetime is evaluated using an average lifetime ($\tau_{ave}$) and corresponding fitting results are shown in Fig. S8 and Table S2. Compared to other amounts of Pb(OAc)$_2$ excess, the 1.2 film shows longer $\tau_{ave}$, consistent with a reduced density of trap states in the 1.2 perovskite film.[56,57] To evaluate the optical properties of the optimal CsPbI$_3$ film (X=1.2) in more detail, the film was characterized using photothermal deflection spectroscopy (PDS), as shown in Fig. S9. The Urbach energy (Eu) extracted from the PDS spectrum is ~18.28 meV - lower than previously reported values for inorganic lead halide perovskites.[58–61] A lower Urbach energy corresponds to a reduced energetic disorder, indicating large diffusion lengths and carrier lifetimes in the X=1.2 films. [35,62,63]

**Photovoltaic performance of γ-CsPbI$_3$ devices with different Pb(OAc)$_2$ excess**

To characterize the photovoltaic performance of the films fabricated using different Pb(OAc)$_2$ excesses, they were integrated into solar cells with the inverted structure of glass/ITO/MeO-2PACz (MeO2)/γ-CsPbI$_3$/PC$_{61}$BM/bathocuproine (BCP)/Ag. We note that we utilized a 1.0 M concentration and 1.0 EDAI$_2$ for the perovskite precursor solution, which were identified as optimal in the preliminary experiments shown in Figs. S10 and S11 in the ESI. The photovoltaic performance parameters of devices with different Pb(OAc)$_2$ excesses are summarized in Fig. 5a-d. Devices with X=1.1 show an average open-circuit voltage (V$_{OC}$) of 0.85 V, which suggests significant non-radiative losses, in agreement with the lowest PL intensity, shortest average carrier lifetime, and poor layer morphology. Gradually increasing the Pb(OAc)$_2$ ratio to 1.2, the V$_{OC}$ is increased, reaching a maximum of 1.24 V. The short-circuit current density (J$_{SC}$) and fill factor (FF) are also dramatically increased, reaching J$_{SC}$s up to 18 mA/cm$^2$ and FFs of 80%. With a further increase in the Pb(OAc)$_2$ excess (X>1.2), the performance is reduced, with significant drops in the V$_{OC}$ and FF. This indicates that samples with too much Pb(OAc)$_2$ excess might exhibit higher recombination loss, resulting in a lower V$_{OC}$ and FF and, consequently, an overall inferior efficiency.[44,64,65]

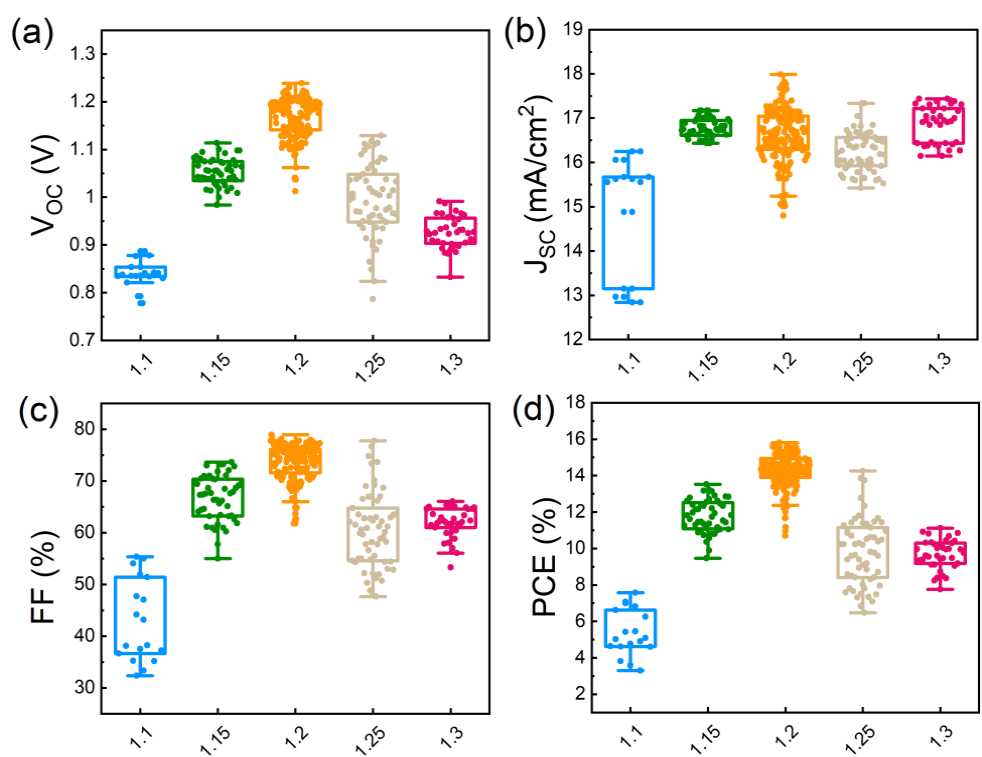

**Fig. 5** Photovoltaic performance parameters: (a) $V_{OC}$, (b) $J_{SC}$, (c) FF and (d) PCE distributions of different Pb(OAc)$_2$ ratio (1.1, 1.15, 1.2, 1.25, 1.3). A total of 350 devices were measured.

**Optimization of hole transport layers for CsPbI$_3$ PSCs**

Considering that the choice of substrates may strongly impact the perovskite film formation,[66–68] we chose the X=1.2 stoichiometry samples to investigate four different hole transport layers (HTL): inorganic (NiOx), polymer (PTAA), self-assemble monolayers (SAMs, MeO-2PACz, in the following termed 'MeO2'), and inorganic/SAMs hybrids (NiOx/MeO-2PACz, NiOx/MeO2 – where the NiOx layer is first deposited on ITO and then coated by a MeO2 SAM). As is detailed in the experimental section, these HTLs were spin-coated on pre-cleaned ITO substrates prior to the deposition of the perovskite layer. To evaluate the impact of the HTL on the perovskite films' microstructure, the layers were characterized by SEM. As is shown in Fig. 6, perovskite layers deposited on NiOx exhibited a multitude of pinholes that could be observed on the surface of the perovskite film, as well as voids that could be seen in the cross-sections. The organic HTLs (PTAA and MeO2) result in better-quality films with no visible pinholes and voids. However, both films exhibit small domains at the film's surface and/or bulk. EDX measurements reveal that these domains are associated with Pb(OAc)$_2$ due to a clear presence of O in these spectra (Figs. S12-14). Finally, the deposition of CsPbI$_3$ on the NiOx/MeO2 hybrid substrates results in a uniform and compact microstructure with no visible pinholes or small Pb(OAc)$_2$ domains at the surface or bulk. To further understand how the choice of HTL influences the morphology of the perovskite layer, we performed contact angle measurements for the different HTLs, with the results shown in Fig. S15. Compared to other HTLs, the PTAA substrate displays the largest contact angle, leading to enlarged perovskite grain size, which is consistent with previous reports that hydrophobic surfaces with increased surface tension can lead to an increased average size of the perovskite domains.[66,69–72] NiOx, MeO2, and NiOx/MeO2 HTLs show almost identical contact angles, thus resulting in a negligible change in the average perovskite grain size. However, there are many pinholes and voids in the perovskite film deposited on NiOx, which is likely related to the presence of surface defects in the NiOx layer.[73,74] To investigate the impact of the HTL on the crystalline structure of the perovskite layer, we performed XRD measurements (Fig. S16). The resultant diffractograms are very similar, with diffraction peaks associated with $Pb(OAc)_2$ and $\gamma$-$CsPbI_3$ and no additional features such as $\delta$-$CsPbI_3$, indicating that the different HTLs do not influence the crystal structure of perovskite even though perovskite morphology changed, which is consistent with previous reports.[69]

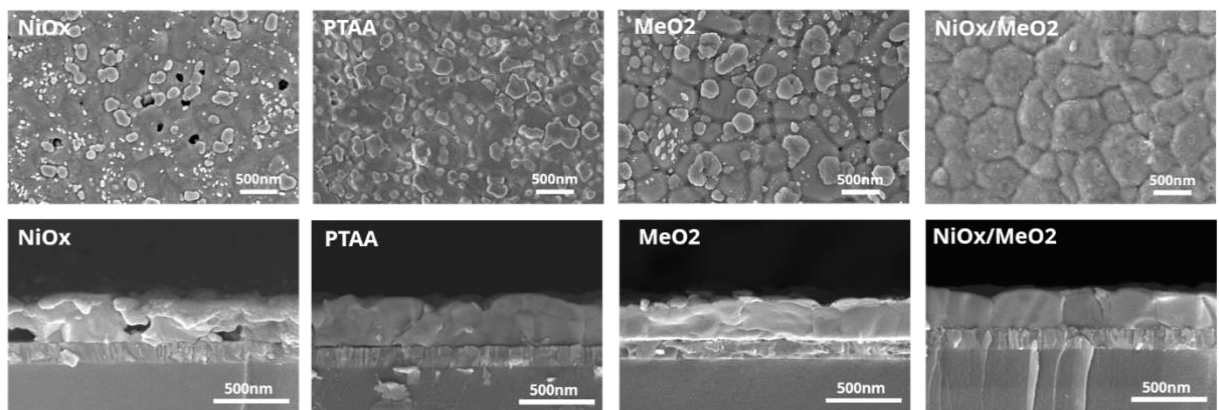

**Fig. 6** Top view and cross-sections SEM images of $CsPbI_3$ films deposited on different hole transport layers (NiOx, PTAA, MeO2, NiOx/MeO2).

To compare the performance of the solar cells with the four different types of HTLs, we fabricated devices with X=1.2 for each of the HTLs. As shown in Fig. 7a-d, devices manufactured on NiOx show the worst performance, particularly in their lower $V_{OC}$s and FFs. This is a direct consequence of the poor film coverage and the many voids at the interface to the HTL. Devices that utilize the PTAA, MeO2, and NiOx/MeO2 HTLs exhibit much better photovoltaic performance. In particular, in the case of the hybrid NiOx/MeO2 HTLs, the devices display much higher $J_{SC}$s and FFs, thus resulting in an overall higher performance. The highest performing device was obtained for NiOx/MeO2 hybrid HTLs (Table 1), achieving an efficiency of 16.57% with a $V_{OC}$ of 1.18 V, $J_{SC}$ of 17.56 mA/cm$^2$, and a FF of 79.74%. To our knowledge, this is among the highest performances reported for $\gamma$-$CsPbI_3$ PSCs with an inverted architecture.[33,34] Importantly, the $CsPbI_3$ PSCs deposited on the different HTLs show negligible hysteresis (Fig. 7e). Fig. 7f displays the external quantum efficiency (EQE) spectrum and integrated $J_{SC}$ curve for the champion $CsPbI_3$ solar cell deposited on NiOx/MeO2. The integrated $J_{SC}$ is calculated to be 17 mA/cm$^2$, which is in excellent agreement with the value extracted from the J-V curve. Figure 7g shows the device's maximum power point (MPP) tracking, resulting in a stabilized power output of 15.60%, which is higher than previously reported values for inverted architecture $\gamma$-$CsPbI_3$ PSCs.[29] A continuous MPP tracking for a period of 12 hours is shown in Fig. S17.

To understand the origin of the enhanced performance of the devices with NiOx/MeO2, we examined their properties in more detail. First, we characterized the morphology of the four different HTLs by atomic force microscopy (AFM). As shown in Fig. S18, the microstructures of NiOx and NiOx/MeO2 are very similar, with essentially the same root mean square (RMS) roughness. The PTAA and MeO2 layers are slightly smoother, but their performance is inferior to that of NiOx/MeO2, suggesting that performance differences are not related to the layer morphology. Next, we examined the energetic alignment at the HTL/perovskite interface and performed ultraviolet photoemission spectroscopy (UPS) measurements on the different HTLs and the perovskite layer, with the results displayed in Fig. S19 and Table S4. It can be seen that among the four extraction layers, only PTAA shows a slight energetic shift as compared to the perovskite layer. In contrast, the other layers show excellent energetic alignment. Despite this energetic offset, PTAA-based devices deliver high $V_{OC}$ averaging 1.17 V, due to the ability of the perovskite layer to compensate for interfacial energetic offsets by forming an electronic dipole.[75] Despite the energetic alignment of NiOx and NiOx/MeO2 being very similar, the former exhibits a far lower $V_{OC}$, suggesting this loss is related to non-radiative recombination at the interface due to the increased density of defects. Indeed, previous literature reports suggest that NiOx exhibits a large density of oxygen vacancies and other traps in the NiOx layer.[76–78]

**Table 1** Photovoltaic performance parameters of champion cells on different HTLs (NiOx, PTAA, MeO2 and NiOx/MeO2).

| Different HTLs | $V_{OC}$ [V] | $J_{SC}$ [mA/cm$^2$] | FF [%] | PCE [%] |
|---|---|---|---|---|
| NiOx-Forward | 1.03 | 16.78 | 62.25 | 10.74 |
| NiOx-Reverse | 1.05 | 16.78 | 63.06 | 11.13 |
| PTAA-Forward | 1.17 | 16.85 | 77.26 | 15.18 |
| PTAA-Reverse | 1.17 | 16.85 | 78.10 | 15.46 |
| MeO2-Forward | 1.18 | 17.15 | 71.50 | 14.50 |
| MeO2-Reverse | 1.20 | 17.15 | 76.60 | 15.81 |
| NiOx/MeO2-Forward | 1.15 | 17.56 | 75.25 | 15.22 |
| NiOx/MeO2-Reverse | 1.18 | 17.56 | 79.74 | 16.57 |

To investigate the trap densities in the perovskite films deposited on the four different HTLs, hole-only devices with the architecture of ITO/HTLs/perovskite/Spiro-OMeTAD/Au were fabricated, with the corresponding J-V curves shown in Fig. S20. The trap densities ($N_t$) can be obtained according to the following equation: $V_{TFL} = eN_tL^2/2\varepsilon\varepsilon_0$, where $V_{TFL}$ is the trap-filling voltage, e is the elementary charge, L is the thickness of the perovskite layer, $\varepsilon_0$ is the relative dielectric constant, and $\varepsilon$ is the vacuum permittivity.[23] The hole trap densities of the X=1.2 CsPbI$_3$ samples on NiOx, PTAA, MeO2, and NiOx/MeO2 are $1.47\times10^{15}$ cm$^{-3}$, $1.16\times10^{15}$ cm$^{-3}$, $1.01\times10^{15}$ cm$^{-3}$ and $4.66\times10^{14}$ cm$^{-3}$, respectively. Among the different HTLs, devices based on the NiOx/MeO2 HTL show a lower trap density in the final perovskite layer, contributing to the higher photovoltaic performance of the NiOx/MeO2 solar cells.

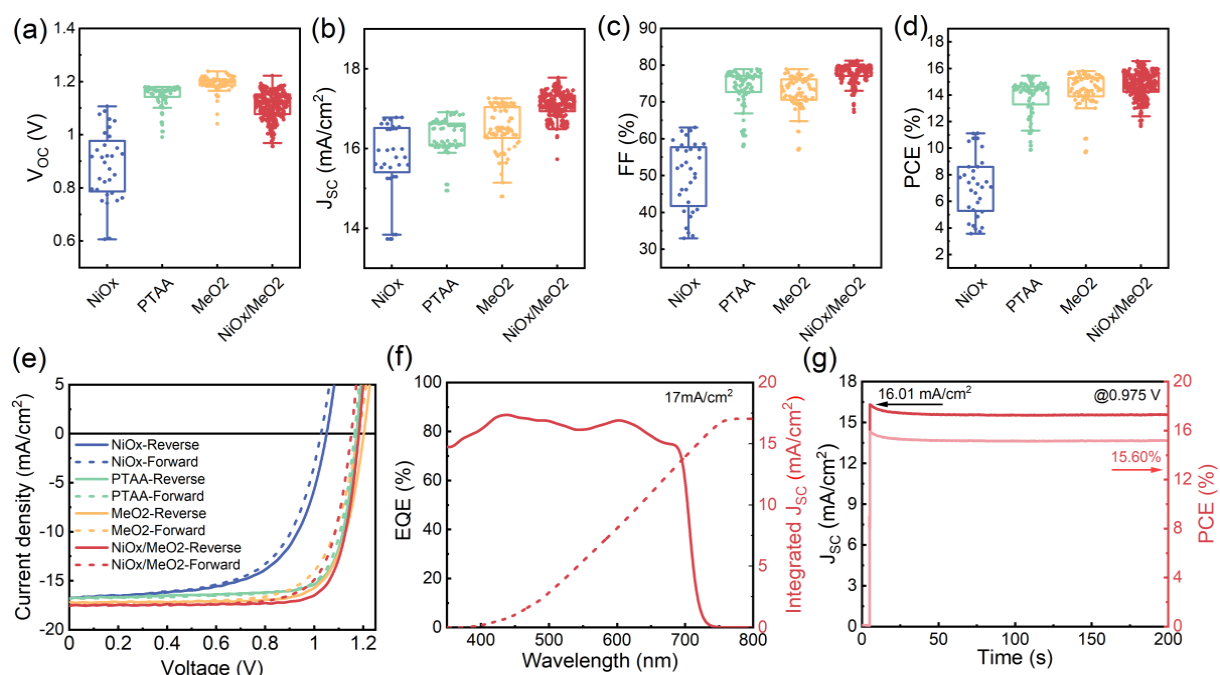

**Fig. 7** Photovoltaic performance parameter: (a) $V_{OC}$, (b) $J_{SC}$, (c) FF, (d) PCE distribution of X=1.2 CsPbI3 PSCs deposited on different HTLs (NiOx, PTAA, MeO2, NiOx/MeO2). A total of 450 devices were measured. e) J-V characteristic curves of champion cells on different HTLs (NiOx, PTAA, MeO2, NiOx/MeO2). (f) EQE spectrum and (g) MPP tracking of 1.2 champion device.

While the new method for fabricating γ-CsPbI$_3$ by a synergetic stabilization via the use of Pb(OAc)$_2$ excess and EDAI$_2$ shows the potential for future tandem photovoltaic applications, it is important to consider the long-term stability of the devices. To gain initial insights into the stability of the devices, we monitored the performance of encapsulated devices that were kept at room temperature under dim illumination (condition is shown in Fig. S21) for a period of 25 days. As shown in Fig. 8a, devices based on a NiOx/MeO2 HTL exhibited the highest stability, maintaining ~97% of their initial PCE for this period of time. Furthermore, we also investigated the light and thermal stability of X=1.2 photovoltaic devices fabricated using the four different HTLs. When exposed to continuous illumination at one sun for 220 hours, the X=1.2 devices maintained ~70% of their initial performance, while devices based on NiOx degraded completely (Fig. 8b). After continuous heating at 80°C for 220 hours, the X=1.2 CsPbI$_3$ devices with a NiOx/MeO2 HTL maintain approximately 55% of their initial efficiency, while the devices based on all other HTLs are completely degraded (Fig. 8c).

To investigate the origin of the different light and thermal stability of the devices based on the various HTLs, SEM and XRD measurements were performed on perovskite films deposited on the different HTLs under continuous one-sun illumination and 80 °C heating at various degradation times. As can be seen in Fig. S22-24, as compared to the other HTLs, perovskite films deposited on NiOx/MeO$_2$ show better film coverage (i.e., no pinholes) and a better retention of the perovskite microstructure after exposure to light and heat. The improved microstructure has been linked to enhanced stability both against environmental factors such as moisture and oxygen as well as illumination.[79–83]

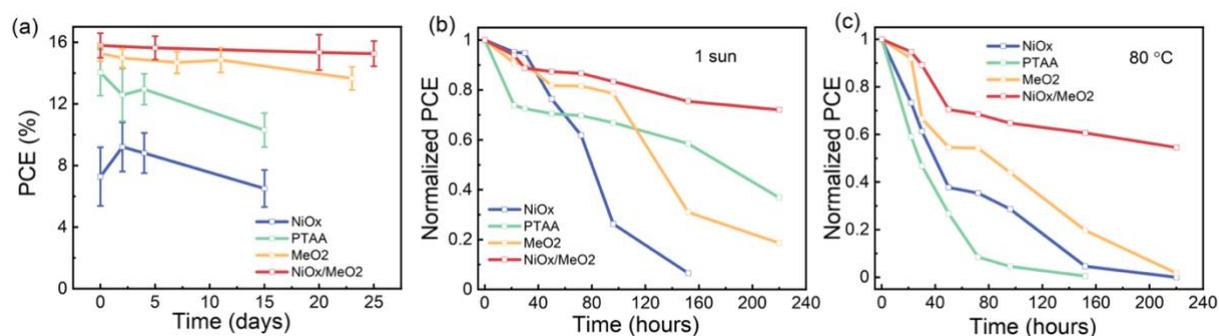

**Fig. 8** Evolution of the PCE of encapsulted devices on different HTLs kept at (a) room temperature, dim light, (b) 1-sun illumination and (c) 80 °C hotplate.

## Conclusions

To summarize, we reported on a new method to fabricate inorganic $CsPbI_3$ perovskites via a low-temperature process that synergistically utilizes $Pb(OAc)_2$ and $EDAI_2$. By adjusting the amount of $Pb(OAc)_2$ excess and selecting the optimal HTL, the microstructure of the perovskite layers could be improved, resulting in high-quality perovskite films with excellent optoelectronic properties. The $EDAI_2:CsI:Pb(OAc)_2=1:1:1.2$ perovskite solar cells reached a high photovoltaic performance of up to 16.57%, which is among the highest previously reported efficiencies for $\gamma$-$CsPbI_3$ inverted PSCs. Moreover, the devices also exhibited enhanced stability and minimal hysteresis. These findings provide a new strategy to stabilize $CsPbI_3$ using a low-temperature process, paving the way for their future application both in tandem configurations and on flexible substrates.

## Experimental Section

### Materials

ITO substrates were obtained from PsiOTech Ltd. $PC_{61}BM$ (>99.5%) was bought from Luminescence Technology Crop. PTAA, NiOx and MeO-2PACz (MeO2) were purchased from Sigma-Aldrich, Liaoning Youxuan Crop and TCI, respectively. Lead (II) Actate (>98%) was obtained from TCI company. Cesium iodide (99.999%, metals basis) was bought from Alfa Aesar. Ethane-1,2-diammonium iodide was purchased from Greatcellsolar Materials. Chlorobenzene, Toluene, Dimethyl sulfoxide, Isopropanol were bought from Acros Organics. All materials were used without more purification.

### Precursor preparation

For PTAA hole transport layer, 1.5mg PTAA powder was dissolved in 1mL toluene solvent then left on the hotplate at 70 °C overnight. For NiOx precursors, 10 mg NiOx powder was dispersed in the 1 mL DI water and ultrasonicated at 40 °C for 10 min, followed filtered with 0.25 um hydrophilic filter. For MeO2 precursors, 1.5mg MeO2 powder was dissolved in the 1 mL isopropanol and followed by ultrasonicated at 40 °C for 10 min. For perovskite precursors, the 259.81 mg (1 mmol) CsI, 315.92 mg (1 mmol) EDAI$_2$ and different ratios of Pb(OAc)$_2$ (357.82 mg, 1.1 mmol; 374.08 mg, 1.15 mmol; 390.35 mg, 1.2 mmol; 406.61 mg, 1.25 mmol; 422.88 mg, 1.3 mmol ) were dissolved in 1 mL pure DMSO and then placed on the hotplate at 70 °C overnight. 20 mg PCBM powder was dissolved in 1 mL chlorobenzene solvent and stirred on hotplate at 70 °C overnight. 0.5 mg BCP powder was dissolved in 1 mL isopropanol and put on hotplate at 70 °C overnight.

**Device fabrication**

Patterned ITO was rinsed with acetone to remove the protective glue. Then they were ultrasonically cleaned with 2 % hellmanex detergent, deionized water, acetone, and isopropanol, followed dry with nitrogen gun. For PTAA hole transport layer, the ITO substrates were treated with oxygen plasma for 10 min and then spin-coated with PTAA at 4000 rpm 30 s, followed by annealing at 100 °C for 10 min in a nitrogen filled glovebox. For nickel oxide, the clean ITO substrates were exposed to UV-ozone for 15 min and then coated with NiOx aqueous at 5000 rpm for 40s and annealed in ambient air at 120 °C for 10 min. For MeO2, the MeO2 solutions was coated on the plasma treated ITO substrates at 3000 rpm for 30s then annealed at 100 °C for 10 min in a nitrogen-filled glovebox. For NiOx/MeO2 hybrid HTLs, the MeO2 was deposited on the NiOx substrates at 3000 rpm for 30s then annealed in glovebox at 100 °C for 10 min. Subsequently, HTLs coated substrates were transferred to the home-made drybox. For PTAA coated layer, in order to increase wettability, 50 ul DMF was spin-coated on substrates at 4000 rpm for 30 s before perovskite deposition. For perovskite active layer, 25 μL precursors were spin-coated on the HTL deposited ITO substrates at 6000 rpm for 60s, and followed by annealing at 60 °C for 5 min then 180 °C for 3 min. Subsequently, the as-prepared perovskite films were transferred to nitrogen filled glovebox for other layer deposition. 35 μL EDAI$_2$ solutions (5 mg/mL in ethanol) were spin-coated on the perovskite layer at 4000 rpm for 30s and followed by annealing at 100 °C for 5 min. Next, 25 μL PC$_{61}$BM solutions was dynamically spin-coated at 2000 rpm for 30 s followed by annealing at 100 °C for 3 min. Finally, 35 μL hole-blocking layer was dynamically deposited on substrates at 4000 rpm for 30 s, followed by 80 nm thermally evaporated Ag cathode (Mantis evaporator, base pressure of $10^{-7}$ mbar). The as-prepared photovoltaic devices were sealed in a glovebox using a transparent clean encapsulation glass, encapsulated by a UV-hardened epoxy glue.

**Photovoltaic Device Characterization**

EQE spectra of the devices were recorded using the monochromatic light of a halogen lamp from 400 nm to 800 nm, the reference spectra were calibrated using the NIST-traceable Si diode (Thorlabs). J-V characteristics were recorded by using a computer controlled Keithley 2450 source measure unit under a solar simulator (Abet Sun 3000 Class AAA solar simulator). The incident light intensity was calibrated via a Si reference cell (NIST traceable, VLSI) and tuned by measuring the spectral mismatch factor between a real solar spectrum, the spectral response of reference cell and perovskite devices. All devices were scanned from short circuit to forward bias (1.3 V) to and reverse with a rate of 0.025 V s$^{-1}$. No treatment was applied prior to measurements. The active area for all devices was 4.5 mm$^2$.

**Scanning-Electron Microscopy (SEM) and energy-dispersive x-ray spectroscopy (EDX)**
A SEM (Gemini 500, (ZEISS, Oberkochen, Germany)) with an acceleration voltage of 3 kV was utilized to obtain the surface and cross-sectional morphology images. For EDX measurements, the Oxford XMaxN-150mm$^2$ detector with a solid angle of roughly 0.05 sr was used. The working distance was 8.5 mm and acceleration voltage was 8 kV.

**X-ray diffraction (XRD)**
XRD patterns were measured in ambient air by using a Bruker Advance D8 diffractometer equipped with a 1.6 kW Cu-Anode (λ = 1.54060 Å) and a LYNXEYE_XE_T 1D-Mode detector. The scans (2theta-Omega mode, 2θ = 5°-30°, step size 0.01°, 0.1 s/step) were measured in Standard Bragg-Brentano Geometry (goniometer radius 420 mm). For grazing incidence XRD, the parameters of scans are 2theta mode, 2θ = 5°-12°, step size 0.01°, 0.1 s/step, The incidence angle (Ω) was fixed at 0.2°, 0.5°, 1°, 2°, respectively.

**UV-vis Absorption and photoluminescence measurements**
A Shimadzu UV-3100 spectrometer was utilized to record the ultraviolet−visible (UV−vis) absorbance spectra. PL measurements were performed using a CW blue laser (405 nm, 20 mW, Coherent) as the excitation source. The PL signal was collected using a NIR spectrometer (OceanOptics). All samples were with encapsulated to prevent the decomposition and enable the all measurements to be were carried out in ambient air at room temperature.

**Time-Correlated Single Photon Counting (TCSPC)**
A TCSPC setup contained of a 375 nm laser diode head (Pico Quant LDHDC375), a PMA Hybrid Detector (PMA Hybrid 40), a TimeHarp platine (all PicoQuant), and a Monochromator SpectraPro HRS-300 (Princeton Instruments). Perovskite films on quartz were excited with the 375 nm laser diode and then the emission was collected by the PMA hybrid detector. The pulse width is ≈44 ps, power is ≈3 mW, the spot size is ≈1 mm$^2$, so the excitation fluence is ≈0.132 J m$^{−2}$. The lifetimes were evaluated using reconvolution algorithms of FluoFit (PicoQuant).

**Photothermal Deflection Spectroscopy (PDS)**
The quartz substrate with perovskite film were mounted in the signal enhancing liquid (Fluorinert FC-770) filled quartz cuvette inside a $N_2$ filled glovebox. Then, the samples were excited using a tunable, chopped, monochromatic light source (150W xenon short arc lamp with a Cornerstone monochromator) and probed using a laser beam (635nm diode laser, Thorlabs) propagating parallel to the surface of the sample. The heat generated through the absorption of light changes the refractive index of the Fluorinert liquid, resulting in the deflection of the laser beam. This deflection was measured using a position sensitive-detector (Thorlabs, PDP90A) and a lock-in amplifier (Amatec SR7230) and is directly correlated to the absorption of the film.

**X-ray photoemission Spectroscopy (XPS) measurement**
The samples were transferred to an ultrahigh vacuum chamber (ESCALAB 250Xi by Thermo Scientific, base pressure: $2 \times 10^{−10}$ mbar) for XPS measurements. XPS measurements were carried out using an XR6 monochromated Al Kα source (hv= 1486.6 eV) and a pass energy of

20 eV. Depth profiling and X=1.2 samples surface etching was performed using an argon gas cluster ion beam with large argon clusters (Ar2000) and an energy of 4 k eV generated by a MAGCIS dual mode ion source. These conditions were found to be optimal for the etching of both organic and perovskite layers.[57,84–86] During XPS depth profiling, the etching spot size was (2.5 × 2.5) mm$^2$ and the XPS measurement spot size was 650 μm. The measurement time per etch level was 8 min.

## Author Contributions

Z.Z. fabricated the perovskite films and devices and performed the SEM, XRD, UV-vis, PL, and PV measurements. R.J. and Q.A. assisted with device fabrication and characterization. Y.H and M.D performed the XPS and UPS measurements and data analysis. Y.L performed the PDS measurement and J.B. evaluated the data. Z.Z. coordinated the project and created the first paper draft. Y.V. supervised the work and revised the manuscript. All authors contributed to the preparation of the manuscript.

## Conflicts of interest

There are no conflicts to declare.

## Acknowledgements

Z. Z. and R.J. are grateful for the financial support by the China Scholarship Council (Scholarship#201806750012 and #201806070145, respectively). Z.Z thanks short-term scholarship from the Graduate Academy of Technische Universität Dresden and the Dresden Center for Nanoanalysis (DCN) for providing access to the SEM measurement. Z.Z also thanks Dr. Tian Luo, Dr. Denys Usov, Dr. Fabian Paulus, Dr. Rongjuan Huang, Dr. Katherina Haase, Dr. Rongjuan Huang, Raquel Dantas Campos and Shaoling Bai for very useful and fruitful discussions. This project has received funding from the European Research Council (ERC) under the European Union's Horizon 2020 research and innovation programme (ERC Grant Agreement n° 714067, ENERGYMAPS) and the Deutsche Forschungsgemeinschaft (DFG) in the framework of the Special Priority Program (SPP 2196) project PERFECT PVs (#424216076).